\documentclass[12pt,a4paper]{article}

\usepackage{amsmath,amsthm,amssymb,amscd,a4wide}

\usepackage{dsfont}

\newcommand{\ud}{\mathrm{d}}
\newcommand{\ui}{\mathrm{i}}
\newcommand{\ue}{\mathrm{e}}
\newcommand{\skl}{\mathrm{S}}
\newcommand{\SO}{\mathrm{SO}}
\newcommand{\SU}{\mathrm{SU}}
\newcommand{\U}{\mathrm{U}}
\newcommand{\Sph}{\mathrm{S}^2}

\newcommand{\al}{\alpha}
\newcommand{\be}{\beta}
\newcommand{\de}{\delta}

\newcommand{\om}{\omega}
\newcommand{\Om}{\Omega}
\newcommand{\ve}{\varepsilon}

\newcommand{\vecA}{\boldsymbol A}

\newcommand{\vecp}{\boldsymbol p}
\newcommand{\vecs}{\boldsymbol s}
\newcommand{\vecx}{\boldsymbol x}
\newcommand{\vecy}{\boldsymbol y}
\newcommand{\vecC}{\boldsymbol C}
\newcommand{\vecal}{\boldsymbol\alpha}
\newcommand{\vecsi}{\boldsymbol\sigma}

\newcommand{\cH}{{\mathcal H}}

\newcommand{\rz}{{\mathbb R}}
\newcommand{\nz}{{\mathbb N}}
\newcommand{\kz}{{\mathbb C}}

\DeclareMathOperator{\mtr}{tr}
\DeclareMathOperator{\vol}{vol}        
\DeclareMathOperator{\op}{op}

\newcommand{\eins}{\mathds{1}}

\numberwithin{equation}{section}

\newtheorem{theorem}{Theorem}[section]

\newtheorem{prop}[theorem]{Proposition}

\begin{document}

\thispagestyle{empty}
\noindent
ULM-TP/02-12\\
December 2002\\

\vspace*{1cm}

\begin{center}

{\LARGE\bf   Semiclassical time evolution and \\
\vspace*{3mm} quantum ergodicity  \\ 
\vspace*{4mm} for Dirac-Hamiltonians\footnote{Contribution to the conference
{\it Theoretical Physics 2002}, Paris, 22-27 July 2002}} \\
\vspace*{3cm}
{\large Jens Bolte}%
\footnote{E-mail address: {\tt jens.bolte@physik.uni-ulm.de}}
{\large and Rainer Glaser}%
\footnote{E-mail address: {\tt rainer.glaser@physik.uni-ulm.de}}

\vspace*{1cm}

Abteilung Theoretische Physik\\
Universit\"at Ulm, Albert-Einstein-Allee 11\\
D-89069 Ulm, Germany 
\end{center}

\vfill

\begin{abstract}
Within the framework of Weyl calculus we establish a quantum-classical 
correspondence for the time evolution of observables generated by a 
Dirac-Hamiltonian. This includes a semiclassical separation of particles 
and antiparticles. We then prove quantum ergodicity for Dirac-Hamiltonians 
under the condition that a skew product of the classical relativistic 
translational motion and relativistic spin precession is ergodic.  
\end{abstract}

\newpage

\section{Introduction}
\label{intro}
Dynamical properties of quantum systems have recently attracted considerable 
attention, in particular in  connection with the question of quantum chaos.
In this context a central problem is to characterise eigenstates of a
quantum Hamiltonian semiclassically in terms of the dynamical behaviour
of the associated classical system. Phase space lifts of eigenfunctions
have proved to be especially suited for such investigations since in
the semiclassical limit they converge to classically invariant measures on 
phase space. This property follows from a dynamical version of the
correspondence principle, which is established in a mathematically rigorous
form through Egorov's theorem \cite{Ego69}. Prominent examples of
invariant measures on phase space are ergodic ones. On the quantum side 
a classically ergodic Liouville (i.e. microcanonical) measure corresponds 
to the property of quantum ergodicity. This concept goes back to Shnirelman
\cite{Shn74} and denotes the semiclassical convergence of the phase
space lifts of almost all quantum eigenfunctions to Liouville measure;
it has been proven in several situations \cite{Zel87,Col85,HelMarRob87}.

Quantum systems that possess spin degrees of freedom in addition to their
translational ones require an extension of Egorov's theorem and of the
concept of quantum ergodicity. In previous studies of non-relativistic 
quantum systems with spin \cite{BolGla00,BolGlaKep01}, whose dynamics are 
generated by a Pauli-type operator, we noticed that in general classical 
ergodicity of the translational motion is not sufficient to guarantee 
quantum ergodicity; the spin dynamics have to be considered as well. Here 
we present an extension of these investigations to Dirac-Hamiltonians. 
The main difference to the non-relativistic situation lies in the 
coexistence of particle and antiparticle states. This requires 
a separation of the Hilbert space into two subspaces that are almost 
invariant under the quantum time evolution. Only in these subspaces the 
usual semiclassical methods can be applied. As a consequence, quantum 
ergodicity for Dirac-Hamiltonians is concerned with projections of 
eigenspinors to the almost invariant subspaces, which in general are only 
approximate eigenspinors of the Hamiltonian. 

Due to lack of space we cannot present proofs here. Together with
further references these can be found in \cite{BolGla02}, where we have
investigated quantum ergodicity for more general matrix valued Hamiltonians.

\section{Background}
\label{sec1}
We consider the Dirac equation
\begin{equation}
\label{Diraceq}
\ui\hbar\frac{\partial}{\partial t} \psi(t,\vecx) = \hat H_D \psi(t,\vecx)
\end{equation}
in a fixed inertial frame in which the potentials $\vecA$ and $\phi$
are time-independent,
\begin{equation}
\label{DiracHam}
\hat H_D = c\vecal\cdot \Bigl( \frac{\hbar}{\ui}\nabla - 
\frac{e}{c}\vecA(\vecx) \Bigr) + mc^2\be + e\phi(\vecx) \ ,
\end{equation}
and the Clifford algebra, realised in the usual Dirac representation, is 
generated by $\vecal$ and $\be$. The propagation of Dirac spinors prescribed 
by (\ref{Diraceq}) therefore takes place in the Hilbert space 
$\cH= L^2(\rz^3)\otimes\kz^4$. 

For the purpose of semiclassical approximations it is useful to represent
quantum observables as Weyl operators. In the case of the Dirac-Hamiltonian 
(\ref{DiracHam}) this means
\begin{equation}
\label{WeylHam}
\bigl( \hat H_D \psi\bigr) (\vecx) = \bigl( \op^W [H_D] \psi\bigr) (\vecx) 
= \frac{1}{(2\pi\hbar)^3}\iint \ue^{\frac{\ui}{\hbar}\vecp\cdot(\vecx-\vecy)}
\,H_D\bigl(\vecp,\tfrac{\vecx+\vecy}{2}\bigr)\,\psi(\vecy)\ \ud y\,\ud p \ .
\end{equation}
Here the Weyl symbol 
$H_D(\vecp,\vecx)\in C^\infty(\rz^3\times\rz^3)\otimes\kz^{4\times 4}$
is the smooth function on the classical phase space $\rz^3\times\rz^3$
that arises from (\ref{DiracHam}) by replacing $\tfrac{\hbar}{\ui}\nabla$ 
with $\vecp$. It hence takes values in the hermitian $4\times 4$ matrices, 
and for each $(\vecp,\vecx)$ possesses the two doubly degenerate 
eigenvalues
\begin{equation}
\label{symbeigen}
h_\pm (\vecp,\vecx) = 
e\phi(\vecx) \pm \sqrt{(c\vecp-e\vecA(\vecx))^2 +m^2c^4} 
\end{equation}
with associated projection matrices $\pi_0^\pm(\vecp,\vecx)$ to the 
$h_\pm(\vecp,\vecx)$-eigenspaces in $\kz^4$. The functions (\ref{symbeigen}) 
can be identified as the classical Hamiltonians of relativistic systems with 
positive or negative energy, corresponding to particles or antiparticles, 
respectively. They generate the Hamiltonian flows $\Phi^t_\pm$ on the phase 
space $\rz^3\times\rz^3$, representing the translational part of the 
classical analogue to the quantum dynamics generated by (\ref{Diraceq}).

It will turn out that the spin part of the relevant classical dynamics is
determined by the equations
\begin{equation}
\label{classprecess}
\dot{\vecs}_\pm(t) = \vecC_\pm(\Phi_\pm^t(\vecp,\vecx)) \times \vecs_\pm(t)
\end{equation}
that describe the Thomas precession of a classical normalised spin 
$\vecs\in\Sph$ along the trajectories $\Phi_\pm^t(\vecp,\vecx)$. Here
$\vecC_\pm(\vecp,\vecx)$ contains the electromagnetic fields and potentials 
as known from the relativistic Thomas precession \cite{Tho27}. For each 
classical Hamiltonian $h_\pm$ the dynamics of the translational and the 
spin degrees of freedom can be combined into a single (non-Hamiltonian, 
skew product) flow 
\begin{equation}
\label{skewprod}
Y_\pm^t (\vecp,\vecx,\vecs) = 
\bigl( \Phi^t_\pm (\vecp,\vecx),\vecs_\pm(t) \bigr) \ , \quad\text{with}
\quad Y_\pm^0 (\vecp,\vecx,\vecs) = (\vecp,\vecx,\vecs) \ ,
\end{equation}
on the combined phase space $\rz^3\times\rz^3\times\Sph$. On 
$\Om_E^\pm\times\Sph$, where $\Om_E^\pm =h_\pm^{-1}(E)\subset\rz^3\times\rz^3$
is an energy shell of $h_\pm$, the flow $Y^t_\pm$ leaves the product measure 
$\ud\mu_E^\pm$ that consists of Liouville measure on $\Om_E^\pm$ and the 
normalised area measure on $\Sph$ invariant. If the energy shells are compact 
we will always assume that their Liouville measures are normalised. Then we
can consider situations in which the flows $Y^t_\pm$ are ergodic with respect 
to the probability measures $\ud\mu_E^\pm$.

There exists an extensive calculus for Weyl operators of the form 
(\ref{WeylHam}) when their symbols obey certain restrictions, see e.g.
\cite{Rob87,DimSjo99}. If, in the present case, $|\vecA(\vecx)|$ is bounded by 
some power $|\vecx|^N$, one can introduce the order function
\begin{equation}
\label{orderfct}
M(\vecp,\vecx) := \sqrt{(c\vecp -e\vecA(\vecx))^2 +m^2c^4} \ .
\end{equation}
It defines the symbol class $\skl(M)$, which consists of all smooth, 
matrix valued functions $B(\vecp,\vecx)$ that satisfy the estimates
\begin{equation}
\label{H_Dsymbol}
\|\partial^\al_p\partial^\be_x B(\vecp,\vecx)\|_{4\times 4} \leq
C_{\al\be}\,M(\vecp,\vecx) \  ,
\end{equation}
where $\|\cdot\|_{4\times 4}$ is some matrix norm. The symbol 
$H_D(\vecp,\vecx)$ of the Dirac-Hamiltonian then is in $\skl(M)$, if the 
potentials $\phi(\vecx)$ and $A_k(\vecx)$ ($k=1,2,3$) are smooth and
$\phi(\vecx)$ as well as all derivatives of $\phi(\vecx)$ and $A_k(\vecx)$
are bounded. In this case the estimate 
\begin{equation}
\label{elliptic}
\| (H_D(\vecp,\vecx)+\ui)^{-1} \|_{4\times 4} \leq C\,M(\vecp,\vecx)^{-1} 
\end{equation}
holds and implies that for small enough $\hbar$ the Dirac-Hamiltonian 
$\hat H_D$ is essentially self-adjoint on the domain 
$C_0^\infty(\rz^3)\otimes\kz^4$. It therefore generates a unitary time 
evolution.

\section{Semiclassical projections}
\label{sec2}
On the level of Weyl symbols there exist the two projection matrices
$\pi_0^\pm(\vecp,\vecx)$ onto the two-dimensional eigenspaces of 
$H_D(\vecp,\vecx)$ in $\kz^4$. These lead to the two classical flows
introduced in the preceeding section that correspond to positive and
negative energies, respectively. In the quantum system a related division
of states into particles and antiparticles would also be desirable. Due to 
the effects of pair creation and annihilation such a separation, however, 
can only possibly be achieved in the semiclassical limit. Weyl quantisation 
suggests $\op^W[\pi_0^\pm]$ as first candidates for the desired quantum 
projectors, but these (bounded) operators in fact are no projectors on the 
Hilbert space $\cH$. One therefore proceeds to an inductive construction
by correcting the symbols order by order in $\hbar$, starting with 
$\pi_0^\pm$ as the lowest order term,
\begin{equation}
\label{projectcorrect}
\hat\Pi^\pm :=\op^W[\pi^\pm] \qquad\text{with}\qquad
\pi^\pm(\vecp,\vecx) \sim \sum_{k=0}^\infty \hbar^k\,\pi^\pm_k(\vecp,\vecx)\ ,
\end{equation}
see \cite{EmmWei96}. The expansion in powers of $\hbar$ has to be understood 
in the sense of an asymptotic series in the symbol class $\skl(1)$ that 
consists of all $B\in C^\infty(\rz^3\times\rz^3)\otimes\kz^{4\times 4}$ 
with bounded derivatives,
$\|\partial_x^\al\partial_p^\be B(\vecp,\vecx)\|_{4\times 4}\leq C_{\al\be}$.
The Weyl quantisation of such a symbol then is a bounded operator on $\cH$
\cite{CalVai71}. The terms in the asymptotic expansion (\ref{projectcorrect}) 
are uniquely fixed by the requirement that (in operator norm)
\begin{equation}
\label{coeffcond}
(\hat\Pi^\pm)^* = \hat\Pi^\pm \ ,\quad
\bigl\| (\hat\Pi^\pm)^2 - \hat\Pi^\pm \bigr\| = O(\hbar^\infty) \ ,\quad 
\bigl\| [\hat H_D,\hat\Pi^\pm] \bigr\| = O(\hbar^\infty) \ .
\end{equation}
That such almost projection operators exist is guaranteed by: 
\begin{prop}
\label{prop:projectex}
Let $A_k$ ($k=1,2,3$) and $\phi$ be smooth potentials such that $\phi$
and all derivatives of $A_k$ and $\phi$ are bounded. Then there exist 
two bounded almost projection operators $\hat\Pi^\pm$ that fulfill 
(\ref{coeffcond}) and that are Weyl quantisations of symbols 
$\pi^\pm\in\skl(1)$ with asymptotic expansions (\ref{projectcorrect}). 
Moreover, the almost projectors provide a semiclassical resolution of 
unity on $\cH$, in the sense that 
$\|\hat\Pi^+ +\hat\Pi^- - \eins_\cH\| = O(\hbar^\infty)$.
\end{prop}
A similar result can be found in \cite{PanSpoTeu02prep}. Following 
\cite{NenSor01} one can now construct genuine projection operators 
$\hat P^\pm$ from $\hat\Pi^\pm$ through the Riesz formula,
\begin{equation}
\label{Rieszproj}
\hat P^\pm := \frac{1}{2\pi\ui}\int_{|z-1|=\tfrac{1}{2}}
\bigl( \hat\Pi^\pm -z \bigr)^{-1}\ \ud z \ .
\end{equation}
These projectors are themselves Weyl operators with symbols in $\skl(1)$ 
and are semiclassically close to the almost projectors, 
$\|\hat P^\pm -\hat\Pi^\pm\| = O(\hbar^\infty)$. However, the operators 
$\hat P^\pm$ in general do not commute with the Dirac-Hamiltonian. They only 
almost commute with $\hat H_D$ in the same way as the operators $\hat\Pi^\pm$. 
Nevertheless, one can relate the semiclassical projectors $\hat P^\pm$ to 
certain spectral projectors of $\hat H_D$: Depending on the behaviour of
the potentials at infinity there exist constants $E_\pm$ such that the 
spectrum of $\hat H_D$ inside $(E_-,E_+)$ is discrete; e.g., if the 
potentials and their derivatives vanish at infinity, $E_\pm=\pm mc^2$. 
For $E\in(E_-,E_+)$ we then introduce the spectral projectors 
$\hat P_{spec}^-$ and $\hat P_{spec}^+$ to the spectral intervals 
$(-\infty,E)$ and $(E,+\infty)$, respectively. 
\begin{prop}
\label{prop:Projectequiv}
If $E\in (E_-,E_+)$ is neither in the spectrum of $\hat H_D$, nor an 
accumulation point thereof, and if the same conditions as in Proposition 
\ref{prop:projectex} hold, the semiclassical projectors are close to the 
spectral projectors, 
$\|\hat P^\pm -\hat P_{spec}^\pm\| = O(\hbar^\infty)$.
\end{prop}
The projectors that almost commute with the Dirac-Hamiltonian allow to
introduce the subspaces $\cH^\pm :=\hat P^\pm\cH$ that in the above
semiclassical sense can be viewed as particle and antiparticle spaces. The
time evolution generated by $\hat H_D$ leaves these subspaces invariant
up to the semiclassically long time scale $T(\hbar)= O(\hbar^{-\infty})$. 

A further consequence of the projectors $\hat P^\pm$ to almost commute
with the Dirac-Hamil\-tonian is that the projections of eigenvectors 
$\psi_n\in\cH$ of $\hat H_D$, $\hat H_D\psi_n =E_n\psi_n$, are quasimodes 
with discrepancies of $O(\hbar^\infty)$. This means that 
$\hat P^\pm\psi_n\in\cH^\pm$ are almost eigenvectors of $\hat H_D$ with 
errors whose $\cH$-norms are of $O(\hbar^\infty)$: 
$\|(\hat H_D -E_n)\hat P^\pm\psi_n\|=O(\hbar^\infty)$. Appropriate phase 
space lifts of the normalised quasimodes 
$\phi^\pm_n := \hat P^\pm\psi_n/\|\hat P^\pm\psi_n\|$ can now be studied
in the semiclassical limit and can be related to the dynamical behaviour
of the respective classical flows $Y^t_\pm$.

\section{Semiclassical time evolution}
\label{sec3}
We now consider the time evolution of observables generated by $\hat H_D$.
For convenience we restrict attention to bounded Weyl operators 
$\hat B =\op^W[B]$, with symbols from the class $\skl(1)$ that possess
an asymptotic expansion in integer powers of $\hbar$, compare 
(\ref{projectcorrect}). We call such operators semiclassical observables.  
The semiclassical projectors $\hat P^\pm$ are of this type and provide a 
natural separation of an observable $\hat B$ into diagonal and off-diagonal 
blocks, $\hat B =\hat B_d +\hat B_{od} +O(\hbar^\infty)$, with
\begin{equation}
\label{Bblocks}
\hat B_d :=\hat P^+\hat B\hat P^+ + \hat P^-\hat B\hat P^- 
\qquad\text{and}\qquad 
\hat B_{od} :=\hat P^+\hat B\hat P^- + \hat P^-\hat B\hat P^+ \ .
\end{equation}
One expects that the diagonal blocks will be propagated semiclassically
by the classical dynamics associated with the eigenvalue functions
$h_+$ and $h_-$, respectively. For the off-diagonal blocks it is not
so obvious how a semiclassical propagation works. In fact, under the
quantum time evolution the off-diagonal blocks will cease to be 
semiclassical observables. A related discussion can be found in \cite{Cor01}.

For a precise statement we require in addition to the assumptions on the 
potentials made previously that their first and all higher derivatives 
are bounded: 
\begin{prop}
\label{invalgebra}
Let $\hat B =\op^W[B]$ be a semiclassical observable. Then for $t>0$ its 
time evolution $\hat B(t)$ generated by the Dirac-Hamiltonian is a 
semiclassical observable, $\hat B(t) =\op^W[B(t)]$ with symbol 
$B(t)\in\skl(1)$ and asymptotic expansion of the type (\ref{projectcorrect}), 
if and only if $\hat B_{od}=O(\hbar^\infty)$. 
\end{prop}
Hence the propagation of the diagonal blocks can be analysed 
semiclassically. To leading order this will happen in terms of the 
Hamiltonian flows $\Phi^t_\pm$ for the translational degrees of freedom 
and the spin-transport matrices $D_\pm\in\SU(2)$ that follow from the 
equation
\begin{equation}
\label{spintranseq}
\dot{D}_\pm(\vecp,\vecx,t)+\tfrac{\ui}{2}\,\vecC_\pm\bigl(
\Phi^t_\pm(\vecp,\vecx)\bigr)\cdot\vecsi\,D_\pm(\vecp,\vecx,t) =0 
\ ,\qquad D_\pm|_{t=0}=\eins_2 \ .
\end{equation}
With suitable isometries  
$V_\pm(\vecp,\vecx):\kz^2\to\pi_0^\pm(\vecp,\vecx)\kz^4$ these spin transport
matrices can be converted to $d_\pm(\vecp,\vecx,t)\in\U(4)$ via
$D_\pm(\vecp,\vecx,t)=V_\pm^*(\Phi^t_\pm(\vecp,\vecx))d_\pm(\vecp,\vecx,t)
V_\pm(\vecp,\vecx)$.
By further introducing the notation 
$\hat B^\pm := \hat P^\pm\hat B\hat P^\pm = \op^W[B^\pm]$ we can now
state the relevant Egorov theorem:
\begin{theorem}
\label{thm:Egorov}
Under the conditions specified above the quantum time evolution of the 
diagonal part $\hat B_d$ of a semiclassical observable $\hat B =\op^W[B]$ 
is again a semiclassical observable, $\hat B_d(t) =\op^W[B_d(t)]$,
with symbol $B_d(t)\in\skl(1)$ and asymptotic expansion 
\begin{equation}
\label{Egorovexpan}
B_d(t)(\vecp,\vecx) \sim \sum_{k=0}^\infty \hbar^k\,
B_d(t)_k(\vecp,\vecx) \ .
\end{equation}
The $\hbar$-independent, leading term is completely determined by
the Hamiltonian flows $\Phi^t_\pm$ generated by the eigenvalue functions
$h_\pm$, and by the unitary spin-transport matrices $d_\pm$,
\begin{equation}
\label{Egorovprincip}
B_d(t)_0(\vecp,\vecx) = \sum_{\nu\in\{+,-\}}d_\nu^{-1}(\vecp,\vecx,t)\,
B^\nu_0\bigl(\Phi^t_\nu(\vecp,\vecx)\bigr)\,d_\nu(\vecp,\vecx,t) \ .
\end{equation}
\end{theorem}
The two types of dynamics that enter on the right-hand side of 
(\ref{Egorovprincip}) can be combined into skew product flows on 
$\rz^3\times\rz^3\times\SU(2)$ given by
\begin{equation}
\label{skewgroup}
\tilde Y_\pm^t (\vecp,\vecx,g) := 
\bigl( \Phi^t_\pm (\vecp,\vecx),D_\pm(\vecp,\vecx,t)g \bigr) \ .
\end{equation}
The double covering map $R:\SU(2)\to\SO(3)$ now allows to relate these
flows to the genuinely classical skew product flows (\ref{skewprod}), if one 
sets $\vecs_\pm(t)=R(D_\pm(\vecp,\vecx,t))\vecs$. Moreover, the  dynamical
properties of the flows $Y_\pm^t$ and $\tilde Y_\pm^t$ are closely related.
E.g., $Y_\pm^t$ is ergodic with respect to the product measure $\ud\mu^\pm_E$
on $\Om_E^\pm\times\Sph$, if and only if $\tilde Y_\pm^t$ is ergodic on
$\Om_E^\pm\times\SU(2)$ with respect to the product of Liouville and
(normalised) Haar measure. Thus, to leading semiclassical order the time 
evolution of block-diagonal observables, generated by the Dirac-Hamiltonian 
$\hat H_D$, can be completely described by the two classical flows $Y^t_\pm$ 
that combine the translational and the spin degrees of freedom of particles 
and antiparticles, respectively.

\section{Quantum ergodicity}
\label{sec4}
In quantum systems without spin quantum ergodicity means that phase 
space lifts of almost all eigenfunctions of the quantum Hamiltonian 
semiclassically converge to Liouville measure on an appropriate energy 
shell, if the classical Hamiltonian flow on this energy shell is ergodic. 
Apart from an Egorov theorem, a proof of this statement requires a 
(Szeg\"o-type) limit theorem for averaged phase space lifts of 
eigenfunctions. 

In the case of a Dirac-Hamiltonian one first has to ensure the very 
existence of eigenspinors $\psi_n\in\cH$. To this end we require that 
there exists an energy $E$ such that all energy shells $\Om_{E'}^\pm$ are
compact when $E'$ varies in $[E-\ve,E+\ve]$ with some $\ve>0$. These
$E'$ shall moreover be no critical values of the eigenvalue functions
(classical Hamiltonians) $h_\pm(\vecp,\vecx)$. In addition, at least one
of the energy shells $\Om_E^\pm$ shall be non-empty. Then, for sufficiently
small $\hbar$, the spectrum of $\hat H_D$ is discrete in the interval 
$I(E,\hbar):=[E-\hbar\om,E+\hbar\om]$, comprising of $N_I >0$ eigenvalues.

The desired limit theorem is concerned with averages of the expectation 
values of a semiclassical observable $\hat B=\op^W[B]$ in normalised 
eigenstates $\psi_n$ of $\hat H_D$ with eigenvalues $E_n\in I(E,\hbar)$. 
For this to hold we require that the periodic orbits of the Hamiltonian 
flows $\Phi^t_\pm$ on $\Om_E^\pm$ are of Liouville measure zero; e.g., this 
condition is fulfilled if the flows are ergodic. 
\begin{prop}
\label{prop:Szegoe}
Under the conditions imposed above on the Dirac-Hamiltonian the number
$N_I$ of eigenvalues in the interval $I(E,\hbar)$ grows semiclassically
according to
\begin{equation}
\label{Weylasympt}
N_I = \frac{2\om}{\pi}\frac{\vol\Om_E^+ +\vol\Om_E^-}{(2\pi\hbar)^2}\,
\bigl(1+O(\hbar)\bigr) \ .
\end{equation}
Moreover, the Szeg\"o-type limit formula,
\begin{equation}
\label{eq:Szegoe}
\lim_{\hbar\to 0}\frac{1}{N_I}\sum_{E_n\in I(E,\hbar)}
\langle\psi_n,\hat B\psi_n\rangle 
= \frac{1}{2}\frac{\sum_{\nu\in\{+,-\}}\vol\Om_E^\nu \mtr
\overline{(\pi_0^\nu B_0 \pi_0^\nu)}^{E,\nu}}{\vol\Om_E^+ +\vol\Om_E^-} \ ,
\end{equation}
holds for any semiclassical observable $\hat B=\op^W[B]$.
\end{prop}
Here $\overline{(\pi_0^\nu B_0 \pi_0^\nu)}^{E,\nu}$ denotes an average 
over $\Om_E^\nu$ with respect to Liouville measure. On average, therefore, 
expectation values of observables in eigenstates of the Hamiltonian 
semiclassically converge to a weighted average of the `classical' projections 
$\pi_0^\nu B_0 \pi_0^\nu$ of the `classical' observable $B_0(\vecp,\vecx)$; 
the weights being determined by the relative volumes of the respective 
energy shells. The factor $\tfrac{1}{2}$ accounts for the two dimensions 
of the spin space (i.e. the range of $\pi_0^\nu$ in $\kz^4$). 
An important consequence of the limit formula (\ref{eq:Szegoe}) is that 
only the diagonal part $\hat B_d$ of an observable gives a non-vanishing 
contribution to the semiclassical average.

To prove quantum ergodicity now requires to combine Szeg\"o-type limits 
with the Egorov-Theorem~\ref{thm:Egorov}. The latter, however, is only 
concerned with block-diagonal observables. We therefore consider only 
observables of the type $\hat P^\nu\hat B\hat P^\nu$. For these 
Proposition~\ref{prop:Szegoe} relates expectation values of $\hat B$ in 
the projected eigenspinors $\hat P^\nu\psi_n$ to the classical weighted 
averages of $\pi_0^\nu B_0 \pi_0^\nu$. At least some of these projected 
eigenspinors can possibly vanish as $\hbar\to 0$. But 
Proposition~\ref{prop:Szegoe} allows to conclude that a positive fraction
of them retains a positive norm in the semiclassical limit; hence these 
can safely be normalised. As discussed in section~\ref{sec2} the 
projected (and normalised) eigenspinors in general only yield quasimodes 
for the Dirac-Hamiltonian with discrepancies $O(\hbar^\infty)$. Thus,
quantum ergodicity in the present context is concerned with quasimodes
rather than with actual eigenstates. The reason for this lies in the
fact that only after the projection can one associate a definite classical 
dynamics to eigenspinors.
\begin{theorem}
\label{thm:QE}
Suppose that all the conditions hold that have previously been imposed on 
the Dirac-Hamiltonian $\hat H_D$, as well as that the skew product flow 
$Y^t_\nu$ ($\nu\in\{+,-\}$ fixed) defined in eq. (\ref{skewprod}) is 
ergodic on $\Om_E^\nu\times\Sph$.
Then in every sequence $\{\phi_n^\nu\}_{n\in\nz}$ of normalised projected 
eigenspinors of $\hat H_D$, with $\|\hat P^\nu\psi_n\|\geq\de$ ($\de>0$
small enough and fixed), there exists a subsequence 
$\{\phi_{n_j}^\nu\}_{j\in\nz}$ of density one, i.e.,
\begin{equation}
\label{density1}
\lim_{\hbar\to 0}
\frac{\#\{j;\ \|\hat P^\nu\psi_{n_j}\|\geq\de,\ E_{n_j}\in I(E,\hbar)\}}
{\#\{n;\ \|\hat P^\nu\psi_n\|\geq\de,\ E_n\in I(E,\hbar)\}} = 1 \ ,
\end{equation}
such that for every semiclassical observable $\hat B=\op^W[B]$,
\begin{equation}
\label{eq:QE}
\lim_{\hbar\to 0}\langle \phi_{n_j}^\nu,\hat B\phi_{n_j}^\nu\rangle =
\tfrac{1}{2}\mtr\overline{(\pi_0^\nu B_0 \pi_0^\nu)}^{E,\nu} \ .
\end{equation}
The subsequence $\{\phi_{n_j}^\nu\}_{j\in\nz}$ can be chosen to be independent
of the observable $\hat B$.
\end{theorem}
The statement of this theorem can be made more transparent, if one chooses
an explicit phase space representation of the quasimodes $\phi_n^\pm$. 
E.g., if one introduces the matrix valued Wigner transform
\begin{equation}
\label{Wignertrans}
W[\psi](\vecp,\vecx) := \int\ue^{-\frac{\ui}{\hbar}\vecp\cdot\vecy}\,
\overline{\psi(\vecx-\tfrac{1}{2}\vecy)}\otimes\psi(\vecx+\tfrac{1}{2}\vecy)
\ \ud y
\end{equation}
of $\psi\in\cH$, the expectation value of $\hat B$ in a state $\phi_n^\nu$
reads
\begin{equation}
\label{Wignerexpect}
\langle \phi_n^\nu,\hat B\phi_n^\nu\rangle = \frac{1}{(2\pi\hbar)^d}
\iint\mtr\bigl( W[\phi_n^\nu](\vecp,\vecx)\,(\pi_0^\nu B_0\pi_0^\nu)
(\vecp,\vecx) \bigr)\ \ud p\,\ud x \,\bigl(1+O(\hbar)\bigr) \ .
\end{equation}
The conclusion (\ref{eq:QE}) of Theorem~\ref{thm:QE} can therefore be 
rephrased in terms of the Wigner transform (\ref{Wignertrans}) as
\begin{equation}
\label{Wignerlim}
\lim_{\hbar\to 0}\frac{1}{(2\pi\hbar)^d}W[\phi^\nu_{n_j}](\vecp,\vecx) = 
\frac{1}{2}\frac{1}{\vol\Om^\nu_E}\de\bigl( h_\nu (\vecp,\vecx)-E\bigl)
\,\eins_2 \ . 
\end{equation}
Here the limit along the density-one subsequence $\{\phi^\nu_{n_j}\}_{j\in\nz}$
has to be understood in a weak sense, namely after integration with
$\pi_0^\nu B_0\pi_0^\nu$. Thus, the Wigner transforms of the quasimodes
converge to a uniform distribution on the energy shell $\Om_E^\nu$. In
the matrix aspect, which represents spin, the result is an equidistribution
of `spin up' and `spin down'. With an appropriate Weyl calculus for spin
the latter can also be rephrased in terms of an equidistribution on the
unit sphere $\Sph$, see \cite{BolGlaKep01,BolGla02}.

In Theorem~\ref{thm:QE} it was supposed that one of the two skew product
flows $Y^t_\nu$ is ergodic; no assumption was made about the complementary
classical system. If, however, the other energy shell at $E$ is empty,
the norms $\|\hat P^\nu\psi_n\|$ must converge to one. Then the statement
of the theorem applies to a density-one subsequence $\{\psi_{n_j}\}_{j\in\nz}$
of all eigenspinors $\psi_n$ with $E_n\in I(E,\hbar)$ in the form
\begin{equation}
\label{QEalt}
\lim_{\hbar\to 0}\langle \psi_{n_j},\hat P^\nu\hat B\hat P^\nu 
\psi_{n_j}\rangle = 
\tfrac{1}{2}\mtr\overline{(\pi_0^\nu B_0 \pi_0^\nu)}^{E,\nu} \ .
\end{equation}
In this case quantum ergodicity is hence concerned with the eigenspinors
themselves. For a Dirac-Hamiltonian this situation is not untypical, since 
both the particle and the antiparticle energy shell to be non-empty at the
same energy requires very strong potentials, with magnitudes comparable to 
the rest energy $mc^2$. But then also the description of a relativistic
quantum system in a single-particle framework begins to become questionable.

\subsection*{Acknowledgment}
We would like to thank the Deutsche Forschungsgemeinschaft for financial
support under contracts no. Ste 241/15-1 and /15-2.

\end{document}